\documentstyle{article}
\begin{document}
\pagestyle{empty}
\rightline{UG-3/92}
\rightline{May 1992}
\vspace{2truecm}
\centerline{\bf  Spacetime Scale-Invariance and the Super p-Brane}
\vspace{2truecm}
\centerline{\bf E.\ Bergshoeff \footnote
{Bitnet address: bergshoeff@hgrrug5}
}
\vspace{.5truecm}
\centerline{Institute for Theoretical Physics}
\centerline{Nijenborgh 4, 9747 AG Groningen}
\centerline{The Netherlands}
\vspace{.5truecm}
\centerline{and}
\vspace{.5truecm}
\centerline{\bf L.A.J.\ London\footnote{
Bitnet address: lajl1@phx.cam.ac.uk} and P.K.\ Townsend
\footnote{Bitnet address: pkt10@phx.cam.ac.uk}
}
\vspace{.5truecm}
\centerline{DAMTP, Silver Street}
\centerline{University of Cambridge}
\centerline{England}
\vspace{2truecm}
\centerline{ABSTRACT}
\vspace{.5truecm}
We generalize to p-dimensional
extended objects
and type II superstrings a recently proposed Green-Schwarz type I superstring
action in which the tension $T$ emerges as an integration constant
of the equations of motion. The action is spacetime scale-invariant
but its equations of motion are equivalent to those of the standard
super p-brane for $T\ne 0$ and the null super p-brane for $T=0$.
We also show that for $p=1$ the action can be written in
``Born-Infeld'' form.

\vfill\eject
\pagestyle{plain}

\noindent{\bf 1. Introduction}

\vspace{.5cm}

The action for a particle of mass $m$ in $d$-dimensional
Minkowski spacetime with coordinates $\{x^m, m=0,1,\dots ,d-1\}$
is

\begin{equation}
\label{eq:S1}
S=\int\! dt \biggl [ {1\over 2e} {\dot x}^m {\dot x}^n\eta_{mn}
-m^2e \biggr ]
\end{equation}
where $e(t)$ is the worldline einbein and $\eta_{mn}$ the
(mostly plus) Minkowski metric.
This action is invariant under
Poincar\'e transformations in the $d$-dimensional target space but not
under scale (or conformal-boost) transformations. However, this lack of
scale invariance may be viewed, from the point of view of a
{\sl massless} particle in a $(d+1)$-dimensional spacetime,
as a consequence of a particular choice of solution of the
equations of motion. To see this, suppose that $y$ is the coordinate of the
extra dimension and write the action as

\begin{equation}
\label{eq:S2}
S=\int\! dt {1\over 2e} \biggl [ {\dot x}^m {\dot x}^n\eta_{mn}
+ {\dot y}^2 \biggr ]\ .
\end{equation}
The $y$ equation of motion is $\partial_t (e^{-1}{\dot y})=0$,
i.e.\ ${\dot y}=me$ for arbitrary mass parameter $m$. The remaining
equations are then the same as those of (1).
This illustrates the fact that
a massive particle can be viewed as a massless one in a higher dimension,
with the mass interpreted as the component of momentum in the
extra dimension. In the quantum theory the mass $m$ is quantized if $y$
is periodic and a choice of $m$ then amounts to a truncation of a Kaluza-Klein
theory. The variable $y$ can in this case be viewed as parametrizing
the fibre of a
$U(1)$ bundle over (d-dimensional) spacetime.

The Nambu-Goto action for a string, or more generally a p-brane, is analogous
to that of the {\sl massive} particle. To bring out this
analogy it is convenient to write the p-brane action in the form

\begin{equation}
\label{eq:S4}
S=\int\! d^{p+1}\xi \biggl \{
{1\over 2V} {\rm det} (\partial_i x^m \partial_j x^n\eta_{mn} ) - T^2V
\biggr \}
\end{equation}
where $\{\xi^i, i=0,1,\dots ,p\}$ are the worldvolume
coordinates, $V(\xi)$ is an independent world-volume
density, and $T$ is the tension (with units of mass/unit p-volume).
As for the massive particle this action is also
{\sl not} scale invariant. It is natural to wonder
what the analogue of (2) is in this case. This question was addressed
in two recent papers [1,2].
In [1] an additional variable, analogous to $y(t)$,
was introduced, with the interpretation as the coordinate of the fibre of a
$U(1)$ bundle over loop superspace [3]
(or its extension to the space of maps
of a p-brane to superspace). In this formulation the tension appears
as an integration constant of the $y(t)$-equation of motion and
can be interpreted as the momentum along the $U(1)$ fibre.
However, the action proposed in [1] is
not local on the worldsheet/worldvolume.
It was shown subsequently [2] for $p=1$
that the
appropriate local generalization of (2)
is an action
containing an independent worldsheet ``electromagnetic'' gauge field.
We may readily generalize this to a p-brane action
containing an independent p-form gauge potential

\begin{equation}
A={1\over p!}d\xi^{i_p}\dots d\xi^{i_1} A_{i_1\dots i_p}
\end{equation}
where the wedge product of differential forms is understood.
Its $(p+1)$-form field-strength is\footnote {
Note that $F_{i_1\dots i_{p+1}} = (p+1)\partial_{[i_1} A_{i_2\dots
i_{p+1}]}$ since we adopt the conventions that, for p-form $P$
and q-form $Q$, $d(PQ)=PdQ+(-)^q(dP)Q$.}

\begin{equation}
F=dA={1\over (p+1)!}d\xi^{i_{p+1}}\dots d\xi^{i_1}F_{i_1\dots i_{p+1}}
\end{equation}
and the corresponding action is

\begin{equation}
\label{eq:S3}
S=\int\! d^{p+1}\xi \ {1\over 2V} \biggl [ {\rm det}
(\partial_ix \cdot \partial_j x) + 4{\tilde F}^2 \biggr ]
\end{equation}
where ${\tilde F}={1\over (p+1)!}
\epsilon^{i_{p+1}\dots i_1}F_{i_1\dots i_{p+1}}$.
The equation for motion for $A_{i_1\dots i_p}$ is $\partial_i(V^{-1}{\tilde F})
=0$. Choosing the solution ${\tilde F}={1\over 2}TV$ one then finds that the
remaining field equations are those of (3). Moreover, the new action
(6)
has the {\sl target space scale invariance}\footnote{
This should not be confused with the  {\sl worldvolume}
scale invariance of certain formulations of the (super)
p-brane \cite{Li1}.}

\begin{equation}
\label{eq:ta}
x^m \rightarrow \lambda x^m \hskip 1truecm
A_{i_1\dots i_p} \rightarrow \lambda^{p+1}A_{i_1\dots i_p}
\hskip 1truecm V \rightarrow \lambda^{2(p+1)}V
\end{equation}
which is broken by the solution ${\tilde F}={1\over 2}TV$ if $T\ne 0$. This
is entirely analogous to the particle case. In fact, for $p=0$ one has
${\tilde F}={1\over 2}{\dot A}$ and we recover (2)
 on identifying $e=V$ and
$y=A$.
Note that if the ${\tilde F}^2$ term in (6) is omitted we
have the action of the  {\sl null} p-brane [4]. The action (6) can therefore be
 viewed
as a kind of ``higher-dimensional'' extension of the null
p-brane, just as for $p=0$ it is a higher-dimensional
massless particle.

Consider now the supersymmetric extension of these ideas. For
example, the action for the $d=9$ massive superparticle is

\begin{equation}
\label{eq:S6}
S=\int\! dt\ \biggl [ {1\over 2e} \omega\cdot\omega-m^2e+
m\bar\theta{\dot \theta}\biggr ]
\end{equation}
where $\omega^m={\dot x}^m-i\bar\theta\Gamma^m{\dot \theta}$
and $\bar\theta = \theta^TC$ where $C$ is the {\sl symmetric}
charge conjugation matrix. Note that the last term in
(\ref{eq:S6}) is {\sl not} manifestly supersymmetric and can be
interpreted as a Wess-Zumino term. As a consequence of this term,
the mass $m$ appears as a central charge in the supersymmetry algebra
\cite{Az2}.
Central charges have a natural interpretation as components of momentum
in ``extra'' dimensions, and this suggests that it should be possible
to derive the action of the nine-dimensional masssive superparticle
from the action of the massless superparticle in ten dimensions.
The latter can be written in the form

\begin{equation}
\label{eq:S7}
S=\int\! dt\ {1\over 2e} \biggl [
\omega\cdot\omega + (\omega^9)^2\biggr ]
\end{equation}
where $\omega^9={\dot y}-i\bar\theta\Gamma^9{\dot\theta}$. This is the
supersymmetric extension of (\ref{eq:S2}). The $y$-equation of
motion is $\partial_t(e^{-1}\omega^9)=0$ which has the solution
$\omega^9=me$. As for the bosonic case the remaining equations of
motion are those of (\ref{eq:S6}), but note that the ten-dimensional
action is {\sl manifestly} supersymmetric, as there is no
Wess-Zumino term, and has no dimensionful parameters.

For $p>0$ there is a similar Wess-Zumino term in the standard
super p-brane action and this leads to the appearance of a
p-form topological charge in the supersymmetry algebra which is nonzero
for spacetimes with non-trivial p-cycles \cite{Az3}.
This topological charge is obviously not central with respect to the
d-dimensional Poincar\'e group, but is central with respect to the
{\sl global} symmetry group
of spacetime which, in such cases, is always a proper subgroup of the
$d$-dimensional Poincar\'e group.
These topological charges again suggest the existence of some kind of
``higher-dimensional'' manifestly supersymmetric action,
without dimensionful parameters.
Such an action was given in \cite{Az1} but it contains
variables that are not defined locally on the worldvolume.
{}From the above discussion of the
bosonic case one can guess that a local action with the required
properties may be found by
supersymmetrization of (\ref{eq:S3}). This turns out to be the case.
The resulting action is

\begin{equation}
\label{eq:S8}
S=\int\! d^{p+1}\xi\ {1\over 2V}
\biggl (
g+\Phi^2\biggr )
\end{equation}
where $g={\rm det} (\Pi_i\cdot\Pi_j)$ with $\Pi_i^m=
\partial_ix^m-i\bar\theta\Gamma^m\partial_i\theta$ and $\Phi$
is the dual of a supertranslation-invariant ``modified'' field strength
for a worldvolume p-form gauge potential $A$. As a result of this
modification $A$ acquires a non-trivial supersymmetry transformation.

An action of the form (\ref{eq:S8}) was given in \cite{To1} for the
$N=1$ superstring, where it was derived from a free-differential
algebra extension of the supertranslation algebra. In this paper we consider
the general $p$ case and type II superstrings, and we discuss some features of
this formulation not mentioned previously, such as scale invariance.
We hope to persuade the reader that the new scale-invariant
formulation of the super p-brane action presented here is
a natural one.
This is especially true for $p=1$ because the action (\ref{eq:S8})
may in this case be cast in a geometrically suggestive
``Born-Infeld'' form, as we shall show in section 4.
For the $p=0$ case there is the additional
bonus that the $ {\rm massless}$ particle can be quantized
covariantly using twistor methods \cite{Sh1}. One of the motivations
for the work reported here is the hope that,
given a scale-invariant super p-brane action, twistor
methods might again be applicable.
In fact, progress along these lines has recently been
announced \cite{Ga1}.

\vspace{.5truecm}

\noindent {\bf 2. The Free Differential Superalgebra}

\vspace{.5truecm}

We shall begin, as in \cite{To1}, with the Maurer-Cartan equations,

\begin{equation}
\label{eq:Ma}
d\psi =0 \hskip 1.5truecm d\Pi^m-i\bar\psi\Gamma^m\psi=0
\end{equation}
for the $d$-vector  one-form $\Pi^m$ and Grassmann-odd spinor one-form
$\psi$ of the supertranslation algebra. The exterior product of forms
is again
understood in (\ref{eq:Ma}) and in what follows. We now extend this algebra
to a free differential superalgebra \cite{Au1} by the introduction of an
 additional
$(p+1)$-form $F$ subject to

\begin{equation}
\label{eq:dF}
dF+h(\Pi,\psi)=0
\end{equation}
where $h$ is a $ {\rm closed}$ $(p+2)$-form constructed from
$\Pi$ and $\psi$. We choose

\begin{equation}
h= {i\over 2p!}\Pi^{m_p}\dots \Pi^{m_1}\bar\psi\Gamma_{m_1\dots m_p}\psi
\end{equation}
which is closed for the values of $(p,d)$ admitted by the ``branescan''
\cite{Ac1}. For simplicity, we shall assume that $\psi$ is a
Majorana spinor and hence restrict ourselves to $p=1,2$ and $5$, but
this covers most of the interesting cases.

Observe that $h$ cannot be written as $h=db$ if we require that
$b$ be constructed from $\Pi^m$ and $\psi$. This means that
$h$ represents a {\sl non-trivial} class of the
$(p+2)$-th equivariant cohomology group of the supertranslation algebra
\cite{Az4}.
As a consequence it is not possible to set $F=F'+dK$ in such a way
that (\ref{eq:dF}) reduces to $dF'=0$. The free differential superalgebra
defined
by (\ref{eq:Ma}) and (\ref{eq:dF})
is therefore a nontrivial extension of the supertranslation
algebra.

Eqs.\ (\ref{eq:Ma}) and (\ref{eq:dF}) may be solved as follows
in terms of the 0-forms
$Z^M=(x^m,\theta^\alpha)$ and a p-form $A$,
which may be viewed as the coordinates of the ``group manifold''
${\tilde \Sigma}$ associated with the free differential
superalgebra and extending the supertranslation group manifold $\Sigma$:

\begin{equation}
\label{eq:fo}
\psi=d\theta \hskip 1truecm \Pi^m=dx^m-i\bar\theta\Gamma^md\theta
\hskip 1truecm F=dA-b\ .
\end{equation}
Here $b$ is a potential for $h$, i.e.\ $h=db$. From the remarks
above it should be clear that $b$ cannot be written entirely in terms
of $\psi$ and $\Pi^m$ but must involve $x^m$ and/or $\theta$ explicitly.
In fact, one can always arrange for $x^m$ to appear as $dx^m$ at the
cost of undifferentiated $\theta$'s. For such a choice $b$ will be
translation, but not supersymmetry, invariant.
This lack of supersymmetry invariance of $b$ is restricted by the fact that
$db$ {\sl is} invariant, from which it follows that
$\delta_\epsilon b=d(i\bar\epsilon\Delta)$ for some
$p$-form $\Delta$. The (modified) field-strength $F$ will then be invariant
if we ascribe to $A$ the supersymmetry variation

\begin{equation}
\delta_\epsilon A=i\bar\epsilon\Delta\ .
\end{equation}
To find $\Delta$ we observe that, for an arbitrary variation
$\delta Z^M$,

\begin{eqnarray}
\label{eq:db}
\delta b &=&
 d\biggl ( {1\over p!}\Pi^{M_{p+1}}\dots\Pi^{M_2}
(\delta Z^{M_1})b_{M_1\dots M_{p+1}} \biggr )
+
{1\over (p+1)!}\Pi^{M_{p+1}}\dots\Pi^{M_1}
\delta Z^N h_{NM_1\dots M_{p+1}}
\nonumber\\
&=&
d\biggl ( {1\over p!}\Pi^{A_p}\dots \Pi^{A_1}
(\delta Z)^Bb_{BA_1\dots A_p} \biggr )
\\
&&+{1\over (p+1)!}\Pi^{A_{p+1}}\dots
\Pi^{A_1}(\delta Z)^Bh_{BA_1\dots A_{p+1}}\nonumber
\end{eqnarray}
where $(\delta Z)^A=((\delta x^m+i\delta\bar\theta\Gamma^m\theta)\delta_m^a,
\delta\theta^\alpha)$.
For flat superspace the only non-vanishing component of $h$ is

\begin{equation}
\label{eq:h}
h_{\alpha\beta a_1\dots a_p}=i(\Gamma_{a_1\dots a_p})_{\alpha\beta}
\end{equation}
and for a supersymmetry variation
$(\delta_\epsilon Z)^A=(-2i\bar\theta\Gamma^a\epsilon,\epsilon^\alpha)$.
In this case (\ref{eq:db}) reduces to

\begin{eqnarray}
\label{eq:deb}
\delta_\epsilon b &=&
d\biggl (
{1\over p!}\Pi^{A_p}\dots \Pi^{A_1}
(\delta_\epsilon Z)^Bb_{BA_1\dots A_p}
\biggr )\nonumber\\
&+&{1\over p!}i\Pi^{a_p}\dots \Pi^{a_1}(d\bar\theta\Gamma_{a_1\dots a_p}
\epsilon)+{1\over (p-1)!}\Pi^{a_p}\dots\Pi^{a_2}(
d\bar\theta\Gamma_{a_1a_2\dots a_p}d\theta)(\bar\theta\Gamma^{a_1}
\epsilon)\nonumber\\
&=&d \biggl [
{1\over p!}\Pi^{A_p}\dots \Pi^{A_1}
(\delta_\epsilon Z)^Bb_{BA_1\dots A_p} +
{1\over p!} i \Pi^{a_p}\dots \Pi^{a_1}
(\bar\theta\Gamma_{a_1\dots a_p}\epsilon)\biggr ]\\
&&+ {1\over (p-1)!}\Pi^{a_p}\dots \Pi^{a_2}
\biggl [ (d\bar\theta\Gamma^{a_1}d\theta)(\bar\theta\Gamma_{a_1\dots a_p}
\epsilon)+(d\bar\theta\Gamma_{a_1\dots a_p}d\bar\theta)
(\bar\theta\Gamma^{a_1}\epsilon)\biggr ]\nonumber
\end{eqnarray}
where $d\Pi^a=id\bar\theta\Gamma^ad\theta$ has been used to arrive at the
second equality.
We now need to write the last term on the right-hand side of
(\ref{eq:deb}) as an exact form. The procedure for doing this
makes repeated use of the identity

\begin{equation}
(\Gamma^{a_1})_{(\alpha\beta}(\Gamma_{a_1\dots a_p})_{\gamma\delta)}
\end{equation}
which is equivalent to the closure of $h$.
Firstly, this identity implies that

\begin{eqnarray}
(d\bar\theta\Gamma^{a_1}d\theta)(\bar\theta\Gamma_{a_1\dots a_p}\epsilon)
&+&(d
\bar\theta\Gamma_{a_1\dots a_p}d\theta)(\bar\theta\Gamma^{a_1}\epsilon)\\
&=&{2\over 3}d
\biggl [(d\bar\theta\Gamma^{a_1}\theta)(\bar\theta
\Gamma_{a_1\dots a_p}\epsilon)+(d\bar\theta
\Gamma_{a_1\dots a_p}\theta)(\bar\theta\Gamma^{a_1}
\epsilon)\biggr ]\ .\nonumber
\end{eqnarray}
Using this in (\ref{eq:deb}) we obtain

\begin{eqnarray}
\delta_\epsilon b &=& d\biggl \{
{1\over p!}\Pi^{A_p}\dots \Pi^{A_1}(\delta_\epsilon Z)^Bb_{BA_1\dots A_p}
+{1\over p!}i\Pi^{a_p}\dots\Pi^{a_1}
(\bar\theta\Gamma_{a_1\dots a_p}\epsilon)\nonumber\\
&+&{2\over 3(p-1)!}\Pi^{a_p\dots a_2}\biggl [
(d\bar\theta\Gamma^{a_1}\theta)(\bar\theta\Gamma_{a_1a_2\dots a_p}\epsilon)
+(d\bar\theta\Gamma_{a_1a_2\dots a_p}\theta)(\bar\theta\Gamma^{a_1}\epsilon)
\biggr ] \biggr \}\nonumber\\
&+&{2i\over 3(p-1)!}\Pi^{a_p}\dots \Pi^{a_3}
(d\bar\theta\Gamma^{a_2}d\theta)\ \times \\
&&\biggl [
(d\bar\theta\Gamma^{a_1}\theta)(\bar\theta\Gamma_{a_1a_2a_3\dots a_p})
+(d\bar\theta\Gamma_{a_1a_2a_3\dots a_p}\theta)(\bar\theta\Gamma^{a_1}\epsilon)
\biggr ]\ . \nonumber
\end{eqnarray}
For $p=1$ the last term is absent, so the one-form $\Delta$ may be read of from
this expression (it agrees with the result of \cite{To1}).
For $p>1$ the procedure must be continued in order to rewrite this
last term as an exact form. In practice it is simpler to write down the
general form of $\delta_{\varepsilon} b$ as an exact differential with
arbitrary coefficients and then fix them by comparison with (\ref{eq:db}). The
result for $p=2$ may be found in \cite{Az3}. Here we shall give the result for
$p=5$, starting from the following expression for the $6$-form $b$ \cite{Je1}:
\begin{eqnarray}
b & = &
-i(\bar\theta\Gamma_{\mu\nu\rho\sigma\lambda} d\theta)
\biggl [ \Pi^{\mu}\Pi^{\nu}\Pi^{\rho}\Pi^{\sigma}\Pi^{\lambda}
+
i{5\over 2}
\Pi^{\mu}\Pi^{\nu}\Pi^{\rho}\Pi^{\sigma}(\bar\theta\Gamma^{\lambda} d\theta)
\nonumber \\
 & & -
{10\over 3}\Pi^{\mu}\Pi^{\nu}\Pi^{\rho}(\bar\theta\Gamma^{\sigma}
d\theta)(\bar\theta\Gamma^{\lambda} d\theta)
-
i{5\over 2}\Pi^{\mu}\Pi^{\nu}(\bar\theta\Gamma^{\rho} d\theta)
(\bar\theta\Gamma^{\sigma}d\theta)(\bar\theta\Gamma^{\lambda} d\theta)
\nonumber\\
& & +
\Pi^{\mu}(\bar\theta\Gamma^{\nu} d\theta)(\bar\theta\Gamma^{\rho} d\theta)
(\bar\theta\Gamma^{\sigma}d\theta)(\bar\theta\Gamma^{\lambda}
d\theta)\nonumber \\
& & +
i{1\over 6}(\bar\theta\Gamma^{\mu}d\theta)(\bar\theta\Gamma^{\nu}
 d\theta)(\bar\theta\Gamma^{\rho} d\theta)
(\bar\theta\Gamma^{\sigma}d\theta)(\bar\theta\Gamma^{\lambda} d\theta)
\biggr]\ .
\end{eqnarray}
We find that $\delta_{\varepsilon}b=d(\bar\varepsilon\Delta)$ where

\begin{eqnarray}
\bar\varepsilon\Delta
 & = &
i(\bar\varepsilon\Gamma_{\mu\nu\rho\sigma\lambda}\theta)
\Pi^{\mu}\Pi^{\nu}\Pi^{\rho}\Pi^{\sigma}\Pi^{\lambda} -{25\over
6}(\bar\varepsilon\Gamma_{\mu\nu\rho\sigma\lambda}\theta)
(\bar\theta\Gamma^{\lambda}d\theta)\Pi^{\mu}\Pi^{\nu}\Pi^{\rho}\Pi^{\sigma}
\nonumber \\
& & +
{5\over 6}(\bar\theta\Gamma_{\mu\nu\rho\sigma\lambda}d\theta)
(\bar\varepsilon\Gamma^{\lambda}\theta)\Pi^{\mu}\Pi^{\nu}\Pi^{\rho}\Pi^{\sigma}
\nonumber \\
 & & -
i{22\over 3}(\bar\varepsilon\Gamma_{\mu\nu\rho\sigma\lambda}\theta)
(\bar\theta\Gamma^{\sigma}d\theta)(\bar\theta\Gamma^{\lambda}d\theta)
\Pi^{\mu}\Pi^{\nu}\Pi^{\rho}
\nonumber \\
& & -
i{8\over 3}(\bar\theta\Gamma_{\mu\nu\rho\sigma\lambda}d\theta)
(\bar\theta\Gamma^{\sigma}d\theta)(\bar\varepsilon\Gamma^{\lambda}\theta)
\Pi^{\mu}\Pi^{\nu}\Pi^{\rho}\nonumber \\
 & & +
{93\over 14}(\bar\varepsilon\Gamma_{\mu\nu\rho\sigma\lambda}\theta)
(\bar\theta\Gamma^{\rho}d\theta)(\bar\theta\Gamma^{\sigma}d\theta)
(\bar\theta\Gamma^{\lambda}d\theta)\Pi^{\mu}\Pi^{\nu}
\nonumber \\
& & -
{47\over 14}(\bar\theta\Gamma_{\mu\nu\rho\sigma\lambda}d\theta)
(\bar\theta\Gamma^{\rho}d\theta)(\bar\theta\Gamma^{\sigma}d\theta)
(\bar\varepsilon\Gamma^{\lambda}\theta)\Pi^{\mu}\Pi^{\nu}\nonumber \\
 & & +
i{193\over 63}(\bar\varepsilon\Gamma_{\mu\nu\rho\sigma\lambda}\theta)
(\bar\theta\Gamma^{\nu}d\theta)(\bar\theta\Gamma^{\rho}d\theta)
(\bar\theta\Gamma^{\sigma}d\theta)(\bar\theta\Gamma^{\lambda}d\theta)\Pi^{\mu}
\nonumber\\
& & +
i{122\over 63}(\bar\theta\Gamma_{\mu\nu\rho\sigma\lambda}d\theta)
(\bar\theta\Gamma^{\nu}d\theta)(\bar\theta\Gamma^{\rho}d\theta)
(\bar\theta\Gamma^{\sigma}d\theta)
(\bar\varepsilon\Gamma^{\lambda}\theta)\Pi^{\mu}
\nonumber \\
& & -
{793\over 1386}(\bar\varepsilon\Gamma_{\mu\nu\rho\sigma\lambda}\theta)
(\bar\theta\Gamma^{\mu}d\theta)(\bar\theta\Gamma^{\nu}d\theta)
(\bar\theta\Gamma^{\rho}d\theta)(\bar\theta\Gamma^{\sigma}d\theta)
(\bar\theta\Gamma^{\lambda}d\theta)
\nonumber \\
& & +
{593\over 1386}(\bar\theta\Gamma_{\mu\nu\rho\sigma\lambda}d\theta)
(\bar\theta\Gamma^{\mu}d\theta)(\bar\theta\Gamma^{\nu}d\theta)
(\bar\theta\Gamma^{\rho}d\theta)(\bar\theta\Gamma^{\sigma}d\theta)
(\bar\varepsilon\Gamma^{\lambda}\theta).
\end{eqnarray}

\vspace{.5truecm}

\noindent{\bf 3. The Spacetime-Scale-Invariant Super p-Brane}

\vspace{.5truecm}

Let $W$ be the worldvolume of a p-dimensional extended object.
Given a map $f:W\rightarrow {\tilde \Sigma}$, we can pull back the
{\sl supersymmetry invariant} forms
(\ref{eq:fo}) to the world-volume:

\begin{eqnarray}
\label{eq:pb}
f^*(\xi) &=& d\xi^i\partial_i\theta \hskip 1truecm f^*(\Pi^m)=d\xi^i
\Pi_i^m\nonumber\\
f^*(F)&=& {1\over (p+1)!}d\xi^{i_{p+1}}\dots d\xi^{i_1}F_{i_1\dots
i_{p+1}}
\end{eqnarray}
where $\Pi_i^m=\partial_ix^m-i\bar\theta\Gamma^m\partial_i\theta$ and

\begin{equation}
F_{i_1\dots i_{p+1}}=(p+1)\partial_{[i_1}
A_{i_2\dots i_{p+1}]}-b_{i_1\dots i_{p+1}}
\end{equation}
with $b_{i_1\dots i_{p+1}}$ the components of the pull-back $f^*(b)$ of $b$.
We may now construct a manifestly supersymetric worldvolume metric as

\begin{equation}
g_{ij}=\Pi_i^m\Pi_j^n\eta_{mn}\ .
\end{equation}
In addition, the ``modified''
field-strength $F_{i_1\dots i_{p+1}}$ has only one
independent component, which may be written as the (gauge-invariant
and supersymmetric) worldvolume scalar density

\begin{equation}
\Phi={2\over (p+1)!}\epsilon^{i_{p+1}\dots i_1}F_{i_1\dots i_{p+1}}\ .
\end{equation}
By introducing
an independent density $V$ we can now write down the
{\sl manifestly} supersymmetric action

\begin{equation}
\label{eq:S5}
S=\int\! d^{p+1}\xi\, {1\over 2V}(g+ \Phi^2)
\end{equation}
where $g$ is the determinant of $g_{ij}$.

As for the bosonic action (\ref{eq:S3}), this action is invariant under the
target space scale transformations of (\ref{eq:ta}) with

\begin{equation}
\theta \rightarrow \lambda^{1/2}\theta\ .
\end{equation}
We shall see in the following that the equations of motion of our new
action are equivalent to either those of
the standard super p-brane or those of the null super p-brane,
depending on the choice of an integration constant in the $A$
equation of motion.

To obtain the field equations we need the variation of $b_{i_1\dots i_p}$
induced by a general variation $\delta Z^M$ of $Z^M$.
{}From ({\ref{eq:db}) this is

\begin{eqnarray}
{1\over (p+1)!}\epsilon^{i_{p+1}\dots i_1}\delta b_{i_1\dots i_{p+1}}
&=&
{1\over p!}\epsilon^{i_{p+1}\dots i_1}\partial_{i_1}
\biggl [ (\delta Z)^Ab_{Ai_2\dots i_{p+1}}\biggr ]\nonumber\\
&&+{1\over (p+1)!}\epsilon^{i_{p+1}\dots i_1}(\delta Z)^Bh_{Bi_1\dots i_{p+1}}
\end{eqnarray}
where $h_{Bi_1\dots i_{p+1}}=\Pi_{i_{p+1}}^{A_{P+1}}\dots \Pi_{i_1}^{A_1}
h_{BA_1\dots A_{p+1}}$. Using the specific form of $h$
given in (\ref{eq:h}) we find that

\begin{eqnarray}
\label{eq:vb}
{1\over (p+1)!}\epsilon^{i_{p+1}\dots i_1}\delta b_{i_1\dots i_{p+1}}
&=&
{1\over p!}\epsilon^{i_{p+1}\dots i_1}\partial_{i_1}
\biggl [ (\delta Z)^Ab_{Ai_2\dots i_{p+1}}\biggr ]
-{i\over p!}\epsilon^{i_p\dots
i_1j}\partial_j\bar\theta\Gamma_{i_1\dots i_p}
\delta \theta\nonumber\\
&&+{i\over 2(p-1)!}\epsilon^{i_{p-1}\dots i_1jk}\partial_j\bar\theta
\Gamma_a\Gamma_{i_1\dots i_{p-1}}\partial_k\theta(\delta Z)^a\ .
\end{eqnarray}
By defining the matrix

\begin{equation}
\Xi = {1\over (p+1)!}\epsilon^{i_{p+1}\dots i_1}\Gamma_{i_1\dots i_{p+1}}
\end{equation}
which satisfies

\begin{equation}
\Xi^2=-g
\end{equation}
and using the relation

\begin{equation}
\epsilon^{i_{p+1}\dots i_{k+1}i_k\dots i_1}\Gamma_{i_{k+1}\dots
i_{p+1}}=(p-k+1)!\Gamma^{i_k\dots i_1}\Xi
\end{equation}
we can simplify (\ref{eq:vb}) to

\begin{eqnarray}
{1\over (p+1)!}\epsilon^{i_{p+1}\dots i_1}\delta b_{i_1\dots i_{p+1}}
&=&
{1\over p!}\epsilon^{i_{p+1}\dots i_1}\partial_{i_1}
\biggl [ (\delta Z)^Ab_{Ai_2\dots i_{p+1}}\biggr ]\nonumber\\
&+&i\partial_j\bar\theta\Gamma^j\Xi\delta\theta+
{i\over 2}\partial_i\bar\theta\Gamma_a\Gamma^{ij}\Xi\partial_j\theta
(\delta Z)^a\ .
\end{eqnarray}
It is now straightforward to derive
the variation of the action (\ref{eq:S5}) under a general variation
of $Z^M, A_{i_1\dots i_p}$ and $V$:

\begin{eqnarray}
\label{eq:vs}
\delta S &=&
\int d^{p+1}\xi \biggl \{
-{\delta V\over 2V^2}(g+\Phi^2)+2i\partial_i\bar\theta
\Gamma^i({g\over V}-(\Phi V^{-1})\Xi)\delta\theta\nonumber\\
&&-{2\over p!}\epsilon^{i_{p+1}\dots i_1}(\delta A_{i_2\dots i_{p+1}}-
(\delta Z)^Ab_{Ai_2\dots i_{p+1}})\partial_{i_1}(V^{-1}\Phi)\\
&&-(\delta Z)_a \biggl [\partial_i({g\over V}\Pi^{ia})
+(i\Phi V^{-1})\partial_i\bar\theta\Gamma^a\Gamma^{ij}\Xi\partial_j\theta
\biggr ] \biggr \}\ .\nonumber
\end{eqnarray}
{}From this result it can be seen that, like the usual super p-brane action,
the action (\ref{eq:S4}) is invariant under the fermionic gauge transformation

\begin{eqnarray}
\label{eq:k1}
\delta_\kappa x^m &=& -i\delta_\kappa\bar\theta\Gamma^m\theta\hskip
1truecm \delta_\kappa A_{i_1\dots i_p} = (\delta_\kappa Z)^Ab_{Ai_2\dots
i_{p+1}}\\
\label{eq:k2}
\delta_\kappa \theta &=& [(\Phi V^{-1})+V^{-1}\Xi]\kappa\hskip .8truecm
\delta_\kappa V = {4i\over p!}\epsilon^{i_{p+1}\dots i_2i_1}\partial_{i_1}
\bar\theta\Gamma_{i_2\dots i_{p+1}}\kappa
\end{eqnarray}
where $\kappa(\xi)$ is a world-volume scalar but spacetime spinor parameter.
For a transformation of the type (\ref{eq:k1}) only the first two terms in
(\ref{eq:vs})
survive and using (\ref{eq:k2}) and
$\Xi^2=-g$ these are easily seen to cancel.

The $A_{i_1\dots i_p}$ field equation is $\partial_i(V^{-1}\Phi)=0$.
Choosing the solution

\begin{equation}
\Phi = VT
\end{equation}
with $T\ne 0$, the remaining equations reduce to
$V={1\over T}\sqrt {-g}$ and

\begin{equation}
\label{eq:eq}
(1+\Gamma)\Gamma^i\partial_i
\theta =0\nonumber \hskip 1.5truecm
\partial_i(\sqrt {-g}\Pi^{ia})-i\sqrt {-g}\partial_i
\bar\theta\Gamma^{ij}\Gamma\partial_j\theta = 0
\end{equation}
where $\Gamma$ is the matrix

\begin{equation}
\Gamma = {\Sigma\over \sqrt {-g}}
\end{equation}
with the property that $\Gamma^2=1$. Eqs.\ (\ref{eq:eq}) are precisely those
of the standard super p-brane action \cite{Be2}.

If, on the other hand, we choose ${\tilde F}=0$,
then the remaining equations reduce to

\begin{equation}
g=0\hskip 1truecm \partial_i(V^{-1}{\tilde g}^{ij}\Pi_j^a)=0\hskip 1truecm
{\tilde g}^{ij}\Gamma_i\partial_j\theta=0
\end{equation}
where ${\tilde g}^{ij}$ is
the matrix of co-factors of $g_{ij}$.
These are the equations of motion of the null super p-brane which has the
action \cite{Zh1}

\begin{equation}
S=\int\! d^{p+1}\xi\, {1\over 2V}g\ .
\end{equation}
Its $\kappa$-transformations are those of (\ref{eq:k1}) and (\ref{eq:k2})
with $\Phi=0$. We remark that the
null super p-brane action is $\kappa$-invariant for any spacetime
dimension, which shows that $\kappa$-symmetry and spacetime supersymmetry
imply world-volume supersymmetry only if $T\ne 0$.

\vspace{.5truecm}

\noindent {\bf 4. Type II Superstrings}

\vspace{.5truecm}

Among supersymmetric extended object actions, the $p=1$ case is
special because there is the possibility of {\sl extended} (non-minimal)
supersymmetry, i.e.\  the type II Green-Schwarz (GS) superstring\footnote{
It has recently been suggested \cite{Ca1}
that type II 5-branes and type II membranes may also be possible,
by allowing worldvolume fields of spin $>1/2$, but
no $\kappa$-invariant spacetime-Poincar\'e invariant action of this type
has been constructed yet.}.
The spacetime scale-invariant reformulation of the action follows the same
pattern as the general $p$, but type I, case just considered.
However, some of the details differ so we shall now consider this case
separately. We shall also take the opportunity to show how $p=1$ is
special in another respect; both the new type I and type II superstring
actions may be rewritten in ``Born-Infeld'' form.

We start from the $N=2$ superspace closed three-form

\begin{equation}
\label{eq:h2}
h={i\over 2}\Pi^m\biggl [ (d\bar\theta_1\Gamma_m d\theta_1)
-(d\bar\theta_2\Gamma_m d\theta_2)\biggr ]
\end{equation}
where
\begin{equation}
\Pi^m=dx^m-i\bar\theta_1\Gamma^md\theta_1-i\bar\theta_2\Gamma^md\theta_2\ .
\end{equation}
For $d=10$ the minimal spinor is chiral so that two type II actions
are possible according to whether the $d=10$ chirality of the spinors
$\theta_1$ and $\theta_2$ is opposite  (type IIA) or the same
(type IIB) but for the analysis to follow it will not be necessary
to specify the type.

We now introduce the additional two-form $F=dA-b$, as in
(\ref{eq:fo}), where $b$ is a potential for $h$ of
(\ref{eq:h2}); a translation-invariant choice is

\begin{equation}
b=-{i\over 2}dx^m(\bar\theta_1\Gamma_md\theta_1-\bar\theta_2\Gamma_md\theta_2)
+{1\over 2}(\bar\theta_1\Gamma^md\theta_1)(\bar\theta_2\Gamma_md\theta_2)\ .
\end{equation}
Following the analysis of section 3 one can show that under the
$N=2$ supersymmetry transformation

\begin{equation}
\delta\theta_1=\epsilon_1,\hskip .5truecm
\delta\theta_2=\epsilon_2, \hskip .5truecm \delta x^m=
(i\bar\epsilon_1\Gamma^m\theta_1+i\bar\epsilon_2\Gamma^m\theta_2)
\end{equation}
the two-form $b$ acquires the transformation
$\delta_\epsilon b=id(\bar\epsilon_1\Delta_1-\bar\epsilon_2\Delta_2)$ where

\begin{equation}
\Delta_r= -{1\over 2} \biggl (
dx^m +{i\over 3}\bar\theta_r\Gamma^m d\theta_r\biggr ) \Gamma_m\theta_r
\hskip 1.5truecm r=1,2.
\end{equation}
It follows that the ``modified'' field strength
$F$ will be
supertranslation (and Lorentz) invariant provided that we assign to the
independent one-form potential $A$ the supersymmetry transformation

\begin{equation}
\delta A=i\bar\epsilon_1\Delta_1-i\bar\epsilon_2\Delta_2 \ .
\end{equation}
The forms
$(\Pi^m,d\theta_1^\alpha,d\theta_2^\alpha,F)$
can again be viewed as the (left)-invariant
differential forms associated with a free-differential algebra.

{}From these invariant forms we can now construct similar worldsheet
forms with components

\begin{eqnarray}
\Pi_i^m &=& \partial_i x^m-i\bar\theta_1\Gamma^m\partial_i\theta_1
-i\bar\theta_2\Gamma^m\partial_i\theta_2\nonumber\\
(\Pi_i^\alpha)_1 &=& \partial_i\theta_1^\alpha \qquad
(\Pi^\alpha_i)_2 = \partial_i\theta_2^\alpha
\\
F_{ij} &=& \partial_i A_j -\partial_j A_i \nonumber\\
 & & +
 \biggl[{i\over 2}\partial_i x^m
\biggl ( \bar\theta_1\Gamma_m\partial_j\theta_1
-\bar\theta_2\Gamma_m\partial_j\theta_2 \biggr )
-{1\over 2} (
\bar\theta_1\Gamma^m\partial_i\theta_1 )(
\bar\theta_2\Gamma_m\partial_j\theta_2)\hskip .3truecm
-(i\leftrightarrow j)\biggr]\nonumber
\end{eqnarray}
from which we construct the worldsheet metric
$g_{ij}=\sum_{r=1}^2
(\Pi_i^m)_r(\Pi_j^n)_r\eta_{mn}$ and hence the
worldsheet densities ${\sqrt{-{\rm det} g_{ij}}}$ and $\Phi=\epsilon^{ij}
F_{ij}$. The spacetime scale-invariant type II superstring action
may now be written exactly as in (\ref{eq:S5}). However,
for $p=1$, this is equivalent to the
``Born-Infeld-type'' action

\begin{equation}
S=\int\! d^2\xi\ {1\over 2V} {\rm det} \biggl ( g_{ij}+2F_{ij}\biggr )
\end{equation}
since the cross terms between $g_{ij}$ and $F_{ij}$ in the expansion of the
determinant cancel. Following the steps of section 3 one can show that this
action has the $\kappa$-gauge invariance

\begin{eqnarray}
\delta_\kappa x^m &=&
-i\delta_\kappa\bar\theta_1\Gamma^m\theta_1-i\delta_\kappa
\bar\theta_2\Gamma^m\theta_2\nonumber\\
\delta_\kappa A_{i} &=& i\Pi_i^m\biggl ( \bar\theta_1\Gamma_m\delta_\kappa
\theta_1-\bar\theta_2\Gamma_m\delta_\kappa\theta_2\biggr )
+(\bar\theta_1\Gamma^m\partial_j\theta_1)(\bar\theta_2\Gamma_m
\delta_\kappa\theta_2)
-(\bar\theta_2\Gamma^m\partial_j\theta_2)(\bar\theta_1\Gamma_m
\delta_\kappa\theta_1)
\nonumber\\
\delta_\kappa\theta_1 &=&
(g+\epsilon^{ij}F_{ij}\Xi)\kappa_1\hskip .8truecm
\delta_\kappa\theta_2\;\;=\;\; (g-\epsilon^{ij}F_{ij}\Xi)\kappa_2\\
\delta_\kappa V &=& 4iVg\biggl [
(\partial_i\bar\theta_1\Gamma^i\kappa_1)+
(\partial_i\bar\theta_2\Gamma^i\kappa_2)\biggr ]\nonumber
\end{eqnarray}
The type I action in this form is obtained simply by
setting $\theta_2=0$.

\vspace{.5truecm}

\noindent {\bf 5. Comments}

\vspace{.5truecm}

We have emphasized that the new formulation of Green-Schwarz type
 actions is spacetime {\sl scale} invariant. The massless particle
is invariant under the full higher-dimensional conformal group
(including conformal boosts). It is unclear whether there is an
 analogue
of this larger group for $p\ge 1$.
\vfill
 We have concentrated in this paper on flat superspace but the
results are readily generalized to curved space. Indeed, the
action remains that of (\ref{eq:S8}) but now

\begin{equation}
\Pi^A=dZ^ME_M{}^A
\end{equation}
where $E_M{}^A$ is the superspace supervielbein, and

\begin{equation}
F=dA-B
\end{equation}
where $H=dB$ is the supergravity $(p+2)$-form.
We expect that, as usual,
$\kappa$-symmetry will require that the background
supergravity fields satisfy their equations
of motion. An interesting further question
is whether this can be generalised
in a kappa-invariant way to include interactions
with background Yang-Mills fields along the lines of
\cite{De1}.

\vspace{1truecm}

\centerline{\bf Acknowledgements}

\vspace{.5truecm}

For one of us (E.B.) this work has been made possible by a fellowship
of the Royal Netherlands Academy of Arts and Sciences (KNAW).
P.K.T. would like to thank the Institute for Theoretical Physics in
Groningen for its hospitality.

\end{document}